\documentclass[%
 reprint,
 amsmath,amssymb,
 aps,hidelinks=true
]{revtex4-2}

\usepackage{graphicx}
\usepackage{dcolumn}
\usepackage{bm}

\usepackage[english]{babel} 

\begin{document}

\preprint{APS/123-QED}

\title{Unidirectional weak visibility in band gap and singular scattering in conduction band of one-dimensional $\mathcal{PT}$-symmetric photonic crystal}

\author{Tiecheng Wang \textsuperscript{1,2}}
 \email{Corresponding author: tcwang@sxu.edu.cn}
\affiliation{%
 \textsuperscript{1} College of Physics and Electronic Engineering, Shanxi University, 030006, Taiyuan, China\\
 \textsuperscript{2} Collaborative Innovation Center of Extreme Optics, Shanxi University, Taiyuan, 030006, China
}%


\date{\today}

\begin{abstract}
We explore the scattering properties of one-dimensional $\mathcal{PT}$-symmetric photonic crystal. Based on Chebyshev's identity, we derive the mathematical expressions of the transmittance, reflectances and generalized absorptances from both sides of a photonic crystal with $N$ unit cells. Two criteria for differentiating $\mathcal{PT}$-exact phase and $\mathcal{PT}$-broken phase are applied to analyze the scattering properties. In the first criterion originating from Bloch theorem, the complex band structure is studied and the corresponding evolutions of exceptional points, $\mathcal{PT}$-exact phase and $\mathcal{PT}$-broken phase are discussed. Near the exceptional points, singular scattering is found and explained, where transmittance and reflectances from both sides reach very large values simultaneously and tend to infinity, at the same time dramatic amplification from both sides arises. In the band gap, a phenomenon we call unidirectional weak visibility is disclosed and analyzed, where transmittance is zero, the reflectance from one side is very large, while the reflectance from the other side is very small. In the second criterion from eigen-equation of scattering matrix, the corresponding distributions of $\mathcal{PT}$-exact phase and $\mathcal{PT}$-broken phase in band structure are studied. Moreover, the singular scattering, which are related to the poles and zeros of the scattering matrix, and unidirectional weak visibility, which corresponds to the scattering of the eigenstates, are explained thoroughly. After comparing these two criteria, the first criterion is preferable because it is consistent with the universally accepted definition.
\end{abstract}

\maketitle


\section{\label{sec:level1}INTRODUCTION}

In recent years, parity-time-($\mathcal{PT}$)-symmetric photonic structures have attracted much interest \cite{x1,x2}. This is motivated by the study of non-Hermitian quantum mechanics \cite{x3,x4,x5,x6}, the non-Hermitian Hamiltonian is $\mathcal{PT}$-symmetric when the complex potential satisfies ${{V}}^{{*}}\left(-x\right)=V\left(x\right)$. If the eigenstate is also $\mathcal{PT}$-symmetric, the eigenvalue is real and the system is in a $\mathcal{PT}$-exact phase. Otherwise, the eigenstate is not $\mathcal{PT}$-symmetric, two conjugate complex eigenvalues arise and the system is in a $\mathcal{PT}$-broken phase. Similar characteristics to those of non-Hermitian quantum mechanics are found in optics. $\mathcal{PT}$-symmetric photonic systems are characterized by a complex index of refraction with a balanced gain and loss $n^*\left(-x\right)=n\left(x\right)$, have different phases and go through a phase transition from a $\mathcal{PT}$-exact phase to a $\mathcal{PT}$-broken phase at an exceptional point \cite{x7,x8,x9,x10,x11}. At this point, the eigenvalues coalesce and their eigenvectors become parallel.

$\mathcal{PT}$-symmetric photonic structures also exhibit other various intriguing regularities and phenomena in scattering. Through the scattering matrix formalism, the existence of a transition between $\mathcal{PT}$-exact phase and the $\mathcal{PT}$-broken phase has been reported \cite{w21}. Due to the optical reciprocity, the product of the two eigenvalues of the scattering matrix is one \cite{w22}. In a $\mathcal{PT}$-exact phase, both eigenvalues are unimodular, so the corresponding eigenstates exhibits no net amplification nor dissipation; while in a $\mathcal{PT}$-broken phase the unimodularity condition cannot be satisfied, in which case one eigenstate corresponds to amplification, and the other to dissipation \cite{w21}. $\mathcal{PT}$-symmetric photonic structures violate the normal photon flux conservation but obey the generalized unitarity relations; these relations illustrate the existence of anisotropic transmission resonances \cite{w23}. Then these relations are extended from one dimension to higher dimensions, where the interactions between multimode fields are considered \cite{w61}. Moreover, unidirectional invisibility is a typical optical effect for $\mathcal{PT}$-symmetric photonic structures. It has been found that the interplay of Bragg scattering and $\mathcal{PT}$-symmetry allows for unidirectional invisibility in a wide range of frequencies around the Bragg point \cite{w11,w12,w13,w14,w15}, while nonlinear optical structures also support unidirectional invisibility \cite{w16}. A $\mathcal{PT}$-symmetric optical medium can act simultaneously as a laser oscillator and as a coherent perfect absorber \cite{w21,w17,w18,w19,w20,w57}. Direction-dependent $\mathcal{PT}$-phase transition has been observed in a dielectric waveguide structure \cite{w25,w26,w27,w101}. In the forward direction, this transition is thresholdless, whereas in the backward direction it has a nonzero threshold \cite{w24}.

We focus on the scattering properties of $\mathcal{PT}$-symmetric photonic crystal, which is closely related to its band structure. The complex band structure of one dimensional $\mathcal{PT}$-symmetric photonic crystals (1DPTSPCs) has been calculated and analyzed, and it has been found that as the non-Hermiticity increases, two types of $\mathcal{PT}$-phase diagrams occur \cite{w28,w29}. The exceptional contours and band structure in the two dimensional $\mathcal{PT}$-symmetric photonic crystal have also been studied, whose non-Hermitian primitive cell is an integer multiple of the primitive cell of the underlying Hermitian system \cite{w30,w100}. In previous studies, two criteria have been commonly used for differentiating the $\mathcal{PT}$-exact phase and $\mathcal{PT}$-broken phase. One is through the eigen-equation of the effective Hamiltonian \cite{w28,w29,w30,w100} obtained from Bloch theorem, which are revealed through the complex band structure, while the other is through the eigen-equation of the scattering matrix \cite{w21,w22,w23}.

Based on these two criteria, in this paper, we study the transmittances, reflectances, absorptances and emittances properties of 1DPTSPCs comprehensively and thoroughly, furthermore, the criterion which is based on the eigen-equation of effective non-Hermitian Hamiltonian is proved to be preferable. The rest of this paper is organized as follows, in Section \ref{sec:level2}, we review the scattering properties of generic $\mathcal{PT}$-symmetric photonic structures and define the generalized absorptance. In Section \ref{sec:level3}, we present our theoretical model and derive the mathematical expressions of the scattering properties of 1DPTSPCs. In Section \ref{sec:level4}, based on the complex and real band structures and those analytic expressions, we present the scattering spectra of 1DPTSPCs and discuss the scattering properties. In Section \ref{sec:level5}, we apply the scattering matrix formalism to 1DPTSPCs and then explain the scattering phenomena profoundly. In Section \ref{sec:level6}, we compare the two criteria in detail and discuss the uniqueness of criterion. We conclude with a summary in Section \ref{sec:level7}.

\section{\label{sec:level2}GENERALIZED ABSORPTANCE}

\begin{figure}
	\centering
	\includegraphics[width=.48\textwidth]{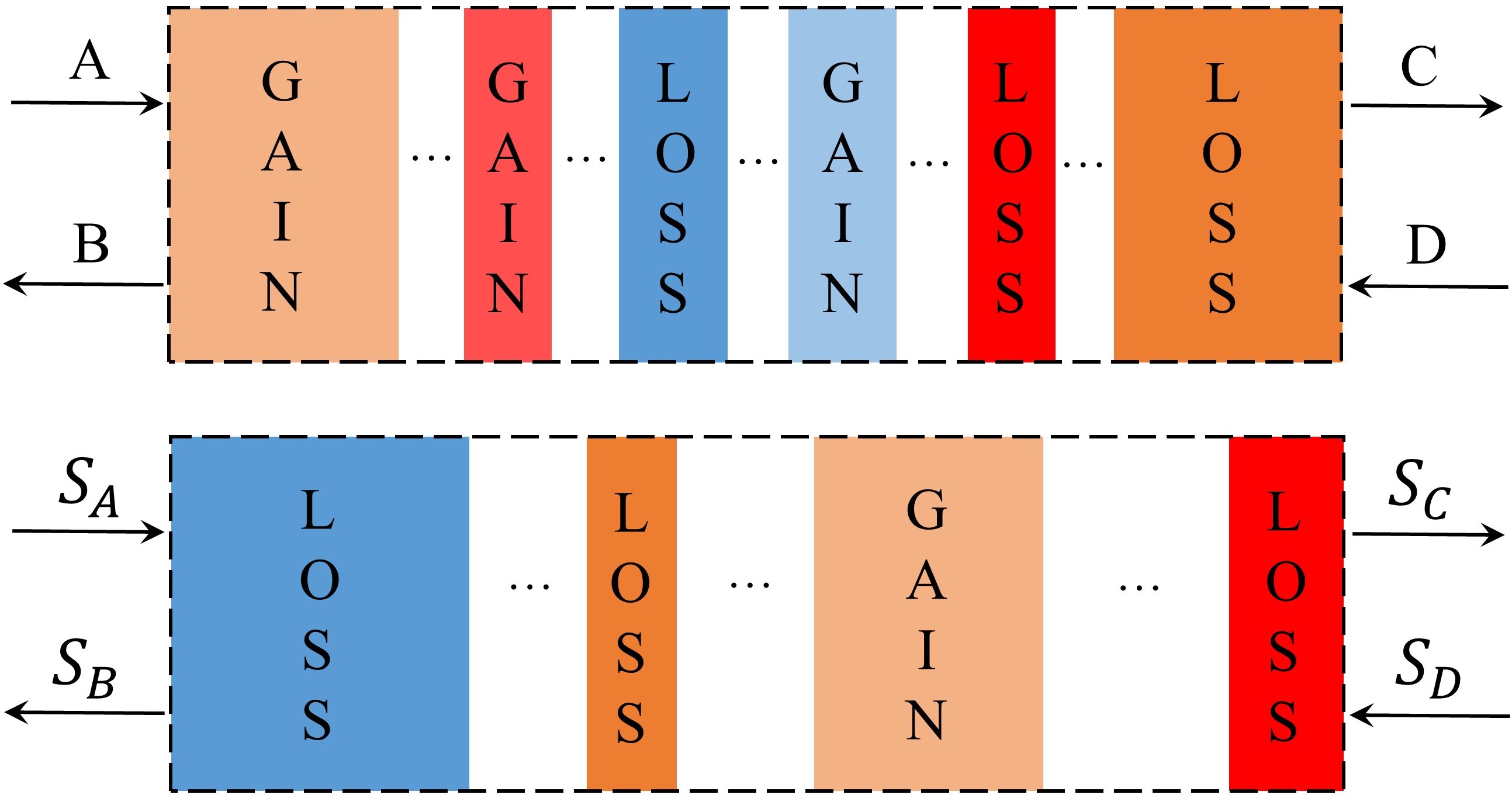}
	\caption{\label{fig:1}(Color online) Schematic pictures of a generic $\mathcal{PT}$ photonic structure (top panel) and a generic optical structure (bottom panel). The $\mathcal{PT}$-symmetric photonic structure is characterized by a complex index of refraction with a balanced gain and loss, while the generic optical structure is made up of dissipative, active or both types of materials.}
\end{figure}

Before we study the transmittances, reflectances and absorptances of 1DPTSPCs, it is worth reviewing the properties of the generic $\mathcal{PT}$-symmetric photonic structure to introduce our discussion. Its transfer matrix $M\left(\omega \right)$, which relates the waves at the two sides of the structure (see the top panel in Fig.~\ref{fig:1}), has been used by many researchers \cite{w31,w32}:
\begin{equation} \label{EQ__1_} 
	\left( \begin{array}{c}
		C \\ 
		D \end{array}
	\right)=\left( \begin{array}{cc}
		M_{11}\left(\omega \right) & M_{12}\left(\omega \right) \\ 
		M_{21}\left(\omega \right) & M_{22}\left(\omega \right) \end{array}
	\right)\left( \begin{array}{c}
		A \\ 
		B \end{array}
	\right),                       
\end{equation} 
where $A$ and $D$ ($B$ and $C$) stand for the amplitudes of the input (output) electromagnetic field from the left and right sides of the generic $\mathcal{PT}$-symmetric photonic structure, respectively. Following the theoretical analysis of \cite{w17}, the components of the transfer matrix $M\left(\omega \right)$ satisfy $M_{22}\left(\omega \right)=M^*_{11}\left({\omega }^*\right)$, $M_{12}\left(\omega \right)=-M^*_{12}\left({\omega }^*\right)$ and $M_{21}\left(\omega \right)=-M^*_{21}\left({\omega }^*\right)$. If the angular frequency $\omega $ is real, the transfer matrix can be parametrized as
\begin{equation} \label{EQ__2_} 
	M\left(\omega \right)=\left( \begin{array}{cc}
		a^* & ic \\ 
		-ib & a \end{array}
	\right),                         
\end{equation} 
where $a$ is complex while $b$ and $c$ are real and related to each other by the condition ${\left|a\right|}^2-bc=1$. The scattering matrix $S\left(\omega \right)$ \cite{w22,w23} can be obtained by transforming the transfer matrix $M\left(\omega \right)$ into the form
\begin{equation} \label{EQ__3_} 
	\left( \begin{array}{c}
		B \\ 
		C \end{array}
	\right)=S\left(\omega \right)\left( \begin{array}{c}
		A \\ 
		D \end{array}
	\right)\equiv \left( \begin{array}{cc}
		r_L & t \\ 
		t & r_R \end{array}
	\right)\left( \begin{array}{c}
		A \\ 
		D \end{array}
	\right),                     
\end{equation} 
where $t\equiv t_L=t_R=1/a$ are the transmission coefficients and $r_L=ib/a$ and $r_R=ic/a$ are the reflection coefficients for the left and right sides, respectively. Based on the relation between those parameters, the generalized conservation law $\left|T-1\right|=\sqrt{R_LR_R}$ can be derived for the transmittance $T={\left|t\right|}^2$ and reflectances $R_L={\left|r_L\right|}^2$, $R_R={\left|r_R\right|}^2$.

If the photonic structure satisfies only parity symmetry ($\mathcal{P}$-symmetry), then $M_{21}\left(\omega \right)=-M_{12}\left(\omega \right)$; if it possesses only time-reversal symmetry ($\mathcal{T}$-symmetry), $M_{22}\left(\omega \right)=M^*_{11}\left({\omega }^*\right)$ and $M_{21}\left(\omega \right)=M^*_{12}\left({\omega }^*\right)$ hold. Consequently, for real angular frequencies, in these two cases the reflectances from the left and right sides are equal, i.e. $R_L=R_R$. This is in contrast to $\mathcal{PT}$-symmetric photonic structures where for real angular frequencies in general $R_L\neq R_R$.

In order to study the scattering properties of 1DPTSPCs as comprehensively as possible, we also consider the absorption or emission of the $\mathcal{PT}$-symmetric structure. To start with, their definitions are as follows. Consider a general optical system, as shown in the bottom panel of Fig.~\ref{fig:1}, where the energy flow densities of the light incident from its left and right sides are denoted by $S_A$ and $S_D$, respectively, and those of the light emitted into the left and right sides are marked by $S_B$ and $S_C$. If this system is made up of absorbing materials with positive imaginary permittivities, the energy flow density which is absorbed is $S_A+S_D-S_B-S_C$. On the other hand, if the imaginary permittivities are negative, the intensity of the light can be amplified, and the increased intensity flow is given by $S_B+S_C-S_A-S_D$. Combining these two cases, if the system is composed of absorbing, active or both types of materials, the generalized absorptance can be defined as
\begin{equation} \label{EQ__7_} 
	W=\frac{S_A+S_D-S_B-S_C}{S_A+S_D}.                             
\end{equation} 
If $W>0$, the optical effect is dissipation; if $W<0$, it is amplification. When $W=0$, the total energy flow density of the outgoing light is equal to that of the incoming light, the optical effect is conservation. 

 When the incoming light is only incident from the left or the right side, the absorptances for these two cases can be respectively expressed as
\begin{equation} \label{EQ__8_} 
	W_L=1-T-R_L, W_R=1-T-R_R.
\end{equation} 

\noindent We can also obtain the relation between $W_L$ and $W_R$ by using the generalized conservation law, as follows: 
\begin{equation} \label{EQ__10_} 
	W_LW_R+\left(T-1\right)\left(W_L+W_R\right)=0.                    
\end{equation} 
At the anisotropic transmission resonances where $T=1$, the incoming lights are unidirectionally reflected. Here the absorption from one side with zero reflectance is zero according to the definitions of $W_L$ and $W_R$. This phenomenon can be also explained using Eq. (\ref{EQ__10_}) directly.

\section{\label{sec:level3}TRANSMITTANCE, REFLECTANCES AND ABSORPTANCES OF 1DPTSPCS}

In our work, through the investigation of the $\mathcal{PT}$-symmetric photonic crystal, we propose that the eigen-equation of the corresponding transfer matrix $M\left(\omega \right)$ are physically meaningful. This eigenvalue spectrum is closely related to the band structure. The complex band structure has already been explored in depth \cite{w28}. We will expound why the $\mathcal{PT}$-symmetric breaking in this criterion is physically meaningful by exploring the scattering properties of 1DPTSPCs.

\begin{figure}
	\includegraphics[width=.45\textwidth]{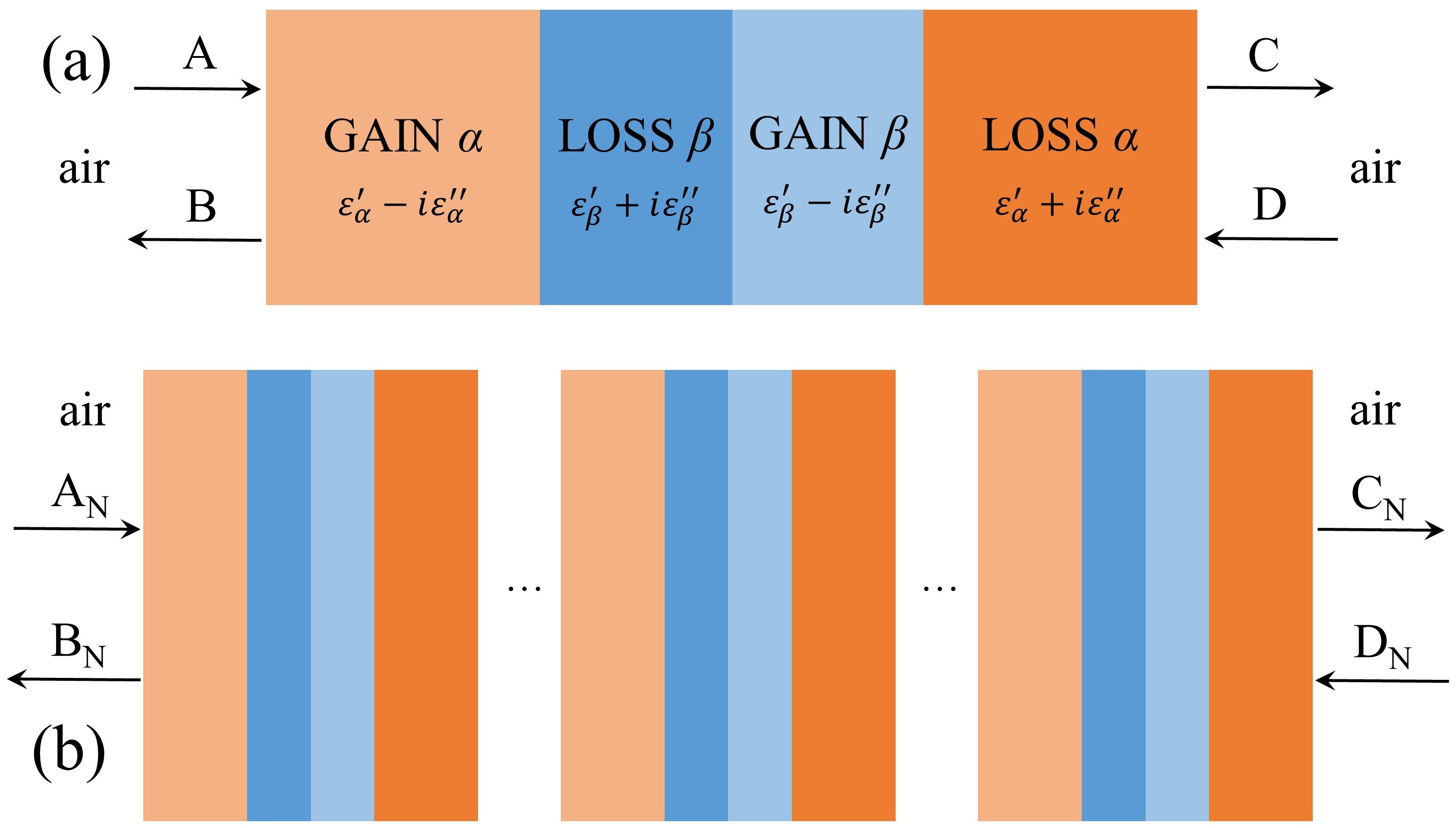}
	\caption{\label{fig:2}{(Color online) Schematic picture of the 1DPTSPC, (a) shows the primitive cell made up of four layers with the relative permittivities }${{\varepsilon }}^{{\prime}}_{{\alpha }}{-}{i}{{\varepsilon }}^{{\prime\prime}}_{{\alpha }}${, }${{\varepsilon }}^{{\prime}}_{{\beta }}{+}{i}{{\varepsilon }}^{{\prime\prime}}_{{\beta }}${, }${{\varepsilon }}^{{\prime}}_{{\beta }}{-}{i}{{\varepsilon }}^{{\prime\prime}}_{{\beta }}${ and }${{\varepsilon }}^{{\prime}}_{{\alpha }}{+}{i}{{\varepsilon }}^{{\prime\prime}}_{{\alpha }}$. (b) is a depiction of the 1DPTSPC with $N$ primitive cells, where $A_N$ and $D_N$ ($B_N$ and $C_N$) are the amplitudes of the input (output) electromagnetic fields from the left and right sides of the 1DPTSPC.}
\end{figure}

We now turn to the study on the 1DPTSPC as shown in Fig.~\ref{fig:2}(b). In our work its primitive cell (see Fig~\ref{fig:2}(a)) is composed of four layers. Their relative permittivities, which describe the balanced gain and loss, are denoted by ${\varepsilon }^{\prime}_{\alpha }-i{\varepsilon }''_{\alpha }$, ${\varepsilon }^{\prime}_{\beta }+i{\varepsilon }''_{\beta }$, ${\varepsilon }^{\prime}_{\beta }-i{\varepsilon }''_{\beta }$ and ${\varepsilon }^{\prime}_{\alpha }+i{\varepsilon }''_{\alpha }$, while their relative permeabilities are equal to one, i.e. $\mu_\alpha=\mu_\beta=1$. The parameters ${\varepsilon }^{\prime}_{\alpha \left(\beta \right)}$ and ${\varepsilon }''_{\alpha \left(\beta \right)}$ are positive real. In the remainder of this work, we use ${\varepsilon }''_{\alpha }={\varepsilon }''_{\beta }={\varepsilon }''$ for simplicity, the real parts of the relative permittivities are fixed at ${\varepsilon }^{\prime}_{\alpha }=1.1$ and ${\varepsilon }^{\prime}_{\beta }=4.0$, and the thicknesses are set to $d_{\alpha }=0.35\Lambda$ and $d_{\beta }=0.15\Lambda$, where $\Lambda$ is the thickness of a primitive cell. Bloch's theorem essentially states the physical meaning of the eigenvalue and eigenfunction of the transfer matrix of a primitive cell
\begin{equation} \label{EQ__11_} 
	\left( \begin{array}{cc}
		M_{11}\left(\omega \right) & M_{12}\left(\omega \right) \\ 
		M_{21}\left(\omega \right) & M_{22}\left(\omega \right) \end{array}
	\right)\left( \begin{array}{c}
		A \\ 
		B \end{array} \right)=e^{iK{\Lambda }}\left( \begin{array}{c}
		A \\ 
		B \end{array}
	\right),                           
\end{equation} 
where $K$ is the Bloch wave vector. Here and in the following, $M(\omega)$ is the transfer matrix of a generic $\mathcal{PT}$-symmetric primitive cell. Based on Bloch's theorem, the real \cite{w40, w41} and complex \cite{w28} band structures can be obtained. The real band structure can be derived by using transfer matrix method ${{\cos} \left(K{\Lambda }\right)\ }={\left(M_{11}\left(\omega \right)+M_{22}\left(\omega \right)\right)}/{2}$. For every real frequency $\omega ={\omega }_a$ we can obtain the corresponding complex wave vector $K=K_a=K_{ar}+iK_{ai}$. If $K_{ai}=0$, the eigenstate is in a conduction band, otherwise it is in a forbidden band. We call this a real band structure because the frequency here is real. In our work, we calculate the complex band structure by using plane wave expansion method which is different from the method in previous research \cite{w28}. Thereupon, for every real wave vector $K=K_c$ we obtain the corresponding complex eigenfrequency $\omega ={\omega }_c={\omega }_{cr}+i{\omega }_{ci}$. In this criterion, if ${\omega }_{ci}=0$, the system is in a $\mathcal{PT}$-exact phase, otherwise it is in a $\mathcal{PT}$-broken phase, in which case the transition boundary corresponds to the exceptional points. We call this a complex band structure because the frequency here is complex. Actually, the whole band structure with complex wave vector and complex frequency is implied in Bloch theorem, the former two types of band structures are only two parts of the whole band structure.

With regard to transmittance and the reflectances from both air sides of the 1DPTSPC with $N$ primitive cells, the total transfer matrix for the model shown in Fig.~\ref{fig:2}(b) is given by
\begin{equation} \label{EQ__12_} 
	\left( \begin{array}{c}
		C_N \\ 
		D_N \end{array}
	\right)={\left( \begin{array}{cc}
			M_{11}\left(\omega \right) & M_{12}\left(\omega \right) \\ 
			M_{21}\left(\omega \right) & M_{22}\left(\omega \right) \end{array}
		\right)}^N\left( \begin{array}{c}
		A_N \\ 
		B_N \end{array}
	\right).                    
\end{equation} 
Here we assume that each primitive cell is surrounded by two infinitely thin films of air on both sides.

In physical scattering, the frequencies of beams are taken to be real. The $N$th power of a unimodular matrix $M\left(\omega \right)$ can be simplified using the Chebyshev's identity \cite{w31}. Using this identity and the parametrized form of Eq.~(\ref{EQ__2_}) for the four-layer primitive cell, we obtain the transmission and reflection coefficients:
\begin{subequations} \label{EQ__14_} 
\begin{equation}
	t_N=\frac{1}{aU_N-U_{N-1}}, 
\end{equation}
\begin{equation}
	r_{NL}{=}\frac{ibU_N}{aU_N-U_{N-1}},\; r_{NR}{=}\frac{icU_N}{aU_N-U_{N-1}},  
\end{equation}           
\end{subequations} 
where $U_N={{\sin} \left( NK_a{\Lambda} \right)}/{{{\sin} \left( K_a{\Lambda } \right)}}$. Furthermore, using the dispersion relation, we have the corresponding transmittance $T_N\equiv {\left|t_N\right|}^2$, reflectances $R_{NL}\equiv {\left|r_{NL}\right|}^2$ and $R_{NR}\equiv {\left|r_{NR}\right|}^{{2}}$, and the absorptances $W_{NL}\equiv 1-T-R_{NL}$ and $W_{NR}\equiv 1-T-R_{NR}$:

\begin{subequations} \label{EQ__15_} 
\begin{equation}
T_N=\frac{1}{bc{{{{\sin}}^{{2}} NK_a{\Lambda }\ }}/{{{{\sin}}^{{2}} K_a{\Lambda }\ }}+1},
\end{equation}

\begin{equation}
R_{NL}=\frac{b^2}{bc+{{{\sin}^2 K_a {\Lambda}\ }}/{{{\sin}^2 NK_a{\Lambda}\ }}},  
\end{equation}

\begin{equation}
R_{NR}{=}\frac{c^2}{bc+{{{{\sin}}^{{2}} K_a{\Lambda }\ }}/{{{{\sin}}^{{2}} NK_a{\Lambda }\ }}},
\end{equation}

\begin{equation}
W_{NL}=\frac{bc-b^2}{bc+{{{{\sin}}^{{2}} K_a{\Lambda }\ }}/{{{{\sin}}^{{2}} NK_a{\Lambda }\ }}},
\end{equation}

\begin{equation}
W_{NR}=\frac{bc-c^2}{bc+{{{{\sin}}^{{2}} K_a{\Lambda }\ }}/{{{{\sin}}^{{2}} NK_a{\Lambda }\ }}}.
\end{equation}
\end{subequations} 

\noindent Because this photonic crystal is composed of $N$ $\mathcal{PT}$-symmetric unit cells, the generalized conservation law also holds for $T_N$, $R_{NL}$ and $R_{NR}$, while the phase relationships between transmission and reflection also hold for $t_N$, $r_{NL}$ and $r_{NR}$. Put another way, the basic generalized unitarity relation $r_{NL}r_{NR}{=}t^2_N\left(1-\frac{1}{T_N}\right)$, which leads to the generalized conservation law, can be proved directly by using the specific expressions of Eq.~(\ref{EQ__14_}).

\begin{figure}
	\centering
	\includegraphics[width=.48\textwidth]{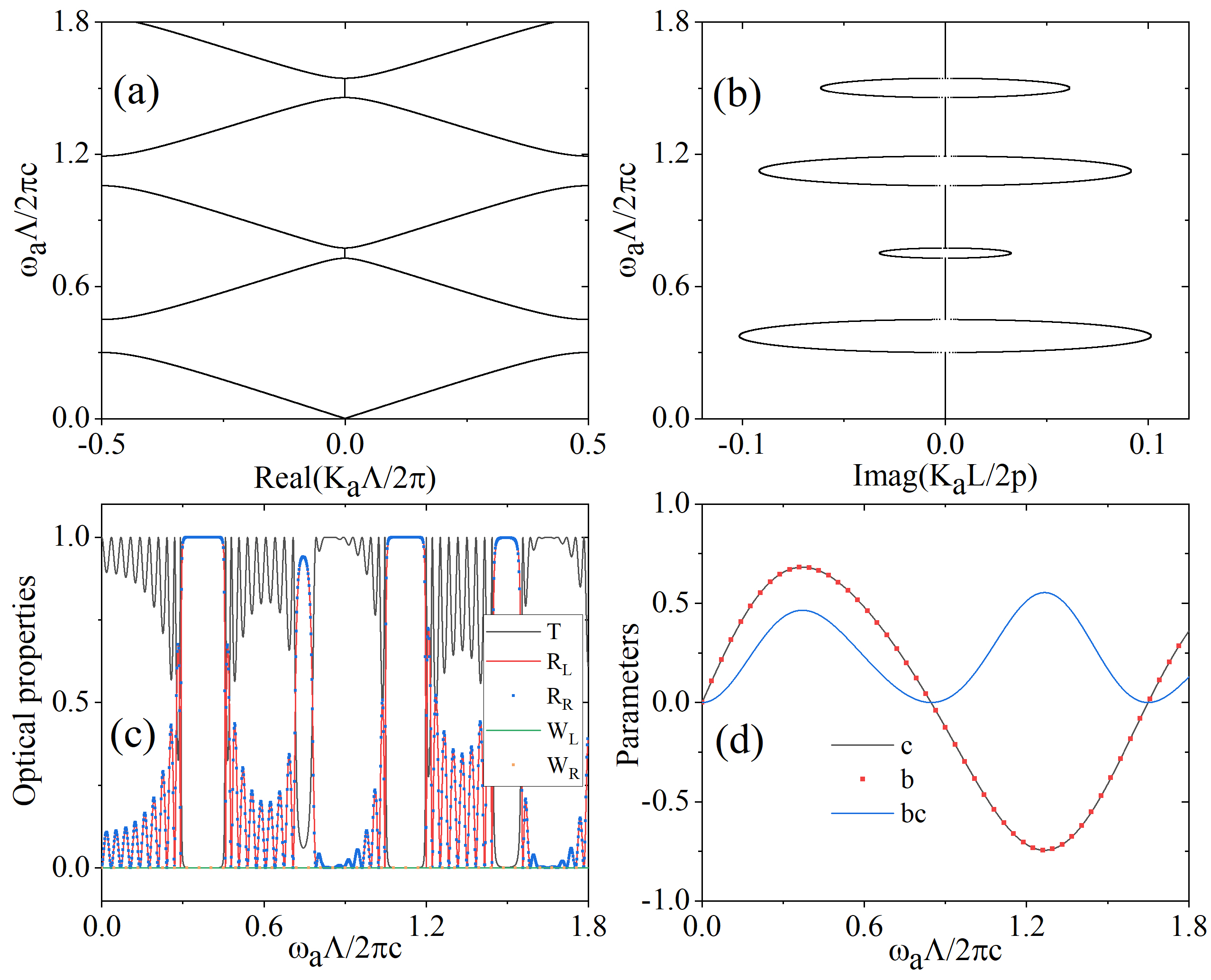}
	\caption{\label{fig:3}(Color online) Real ((a), (b)) band structures of the 1DPTSPC in the absence of gain and loss $(\varepsilon''=0)$. (c) Scattering properties $T_N$, $R_{NL}$, $R_{NR}$, $W_{NL}$ and $W_{NR}$ as functions of the reduced frequency when $\varepsilon''=0$. (d) Values of $c$, $b$ and $bc$ as functions of the reduced frequency when $\varepsilon''=0$. These results are obtained by using transfer matrix method.}
\end{figure}

In the abovementioned scattering properties, if the frequency lies in a conduction band, then, $K_a$ in Eqs.~(\ref{EQ__14_}) and (\ref{EQ__15_}) is a real number, with $K_{ai}=0$. On the other hand, if the frequency lies in a band gap, $K_a{\Lambda}=m\pi +iK_{ai}{\Lambda}$ ($m$ is an integer), and the scattering properties in the band gap can be derived in a similar manner:

\begin{subequations}\label{EQ__17_} 
\begin{equation}
T_N=\frac{1}{bc{{{{\sinh}}^{{2}} NK_{ai}{\Lambda }\ }}/{{{{\sinh}}^{{2}} K_{ai}{\Lambda }\ }}+1},
\end{equation}

\begin{equation}
R_{NL}{=}\frac{b^2}{bc+{{{{\sinh}}^{{2}} K_{ai}{\Lambda }\ }}/{{{{\sinh}}^{{2}} NK_{ai}{\Lambda }\ }}},
\end{equation}

\begin{equation}
R_{NR}=\frac{c^2}{bc+{{{{\sinh}}^{{2}} K_{ai}{\Lambda }\ }}/{{{{\sinh}}^{{2}} NK_{ai}{\Lambda }\ }}},
\end{equation}

\begin{equation}
W_{NL}=\frac{bc-b^2}{bc+{{{{\sinh}}^{{2}} K_{ai}{\Lambda }\ }}/{{{{\sinh}}^{{2}} NK_{ai}{\Lambda }\ }}},
\end{equation}

\begin{equation}
W_{NR}=\frac{bc-c^2}{bc+{{{{\sinh}}^{{2}} K_{ai}{\Lambda }\ }}/{{{{\sinh}}^{{2}} NK_{ai}{\Lambda }\ }}}.
\end{equation}

\end{subequations}

\noindent Compared with the case in Eq.~(\ref{EQ__15_}), these quantities have the same expressions except that the sine function is replaced by the hyperbolic sine function. In a special case, at the band edge, where $K_a{\Lambda}=m\pi$, we have

\begin{figure*}
	
	\includegraphics[width=.98\textwidth]{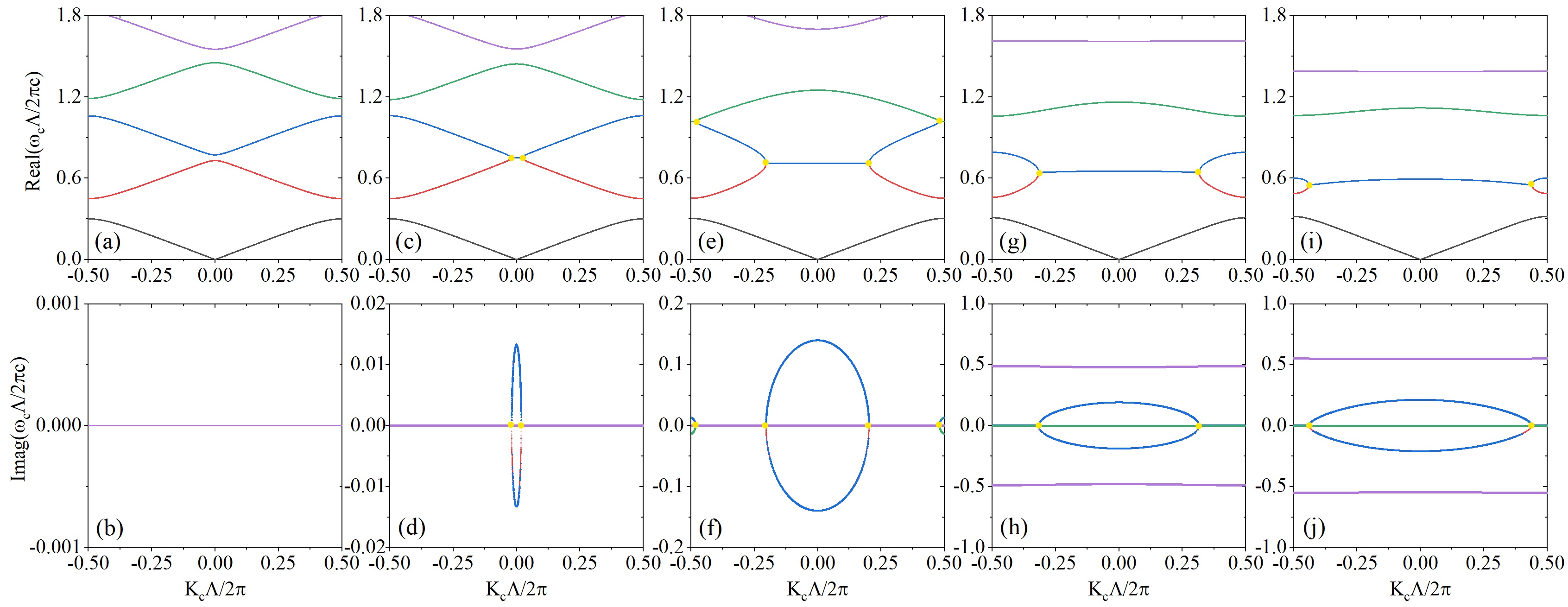}
	
	\caption{\label{fig:4}(Color online) Complex band structures of the 1DPTSPC for different values of $\varepsilon''$. Top graphs: real parts of the complex eigenfrequencies. Bottom graphs: imaginary parts. (a)-(b): $\varepsilon''=0.1$. (c)-(d): $\varepsilon''=0.2$. (e)-(f): $\varepsilon''=1.0$. (g)-(h): $\varepsilon''=1.5$. (i)-(j): $\varepsilon''=2.0$. The black lines denote the eigenfrequencies are real, the other colored lines label complex eigenfrequencies. The yellow dots mark the exceptional points.}
	
\end{figure*}

\begin{subequations}
	
\begin{equation}
T_N=\frac{1}{bcN^2+1},
\end{equation}

\begin{equation}
R_{NL}=\frac{b^2}{bc+{1}/{N^2}}, R_{NR}=\frac{c^2}{bc+{1}/{N^2}},
\end{equation}

\begin{equation}
W_{NL}=\frac{bc-b^2}{bc+{1}/{N^2}}, W_{NR}=\frac{bc-c^2}{bc+{1}/{N^2}}.
\end{equation}

\end{subequations}

So far, we have derived the analytic solutions for the physical quantities $T_N$, $R_{NL}$, $R_{NR}$, $W_{NL}$, and $W_{NR}$ to study the scattering properties of the 1DPTSPC. This theory can be extend to all 1DPTSPCs. 

\section{\label{sec:level4}$\mathcal{PT}$-SYMMETRY BREAKING DEFINED BY EFFECTIVE HAMILTONIAN AND SCATTERING SPECTRA OF 1DPTSPCS}

In this section, we discuss in detail the scattering properties of 1DPTSPCs based on an analysis of their band structures. Figure \ref{fig:3} illustrates the real band structure and corresponding scattering properties in the absence of gain and loss at normal incidence, the number of unit cells is chosen to be $N=10$. We emphasize that the 1DPTSPC here is also $\mathcal{T}$-symmetric, so $b=c$, $R_{NL}=R_{NR}\ge0$ and $W_{NL}=W_{NR}=0$, as derived from Eqs.~(\ref{EQ__15_}) and (\ref{EQ__17_}). For every conduction band, $K{\Lambda}$ varies from $m\pi$ to $(m+1)\pi$. There are $N-1=9$ points where transmission resonances occur. When $K{\Lambda}=n\pi / N$ $\left(n=1,\ 2,\cdots N-1\right)$, so $T_N=1,R_{NL}=R_{NR}=0$ and $W_{NL}=W_{NR}=0$ can obtained from Eq.~(\ref{EQ__15_}). In the band gap, ${{{{\sinh}}^{{2}} NK_{ai}{\Lambda }}}/{{{{\sinh}}^{{2}} K_{ai}{\Lambda }}}$ is very large, so $T_N=0$ can be obtained from Eq.~(\ref{EQ__17_}a), while $R_{NL}=R_{NR}=1$ can be obtained from Eqs.~(\ref{EQ__17_}b) and (\ref{EQ__17_}c).

\begin{figure*}
	
	\includegraphics[width=.98\textwidth]{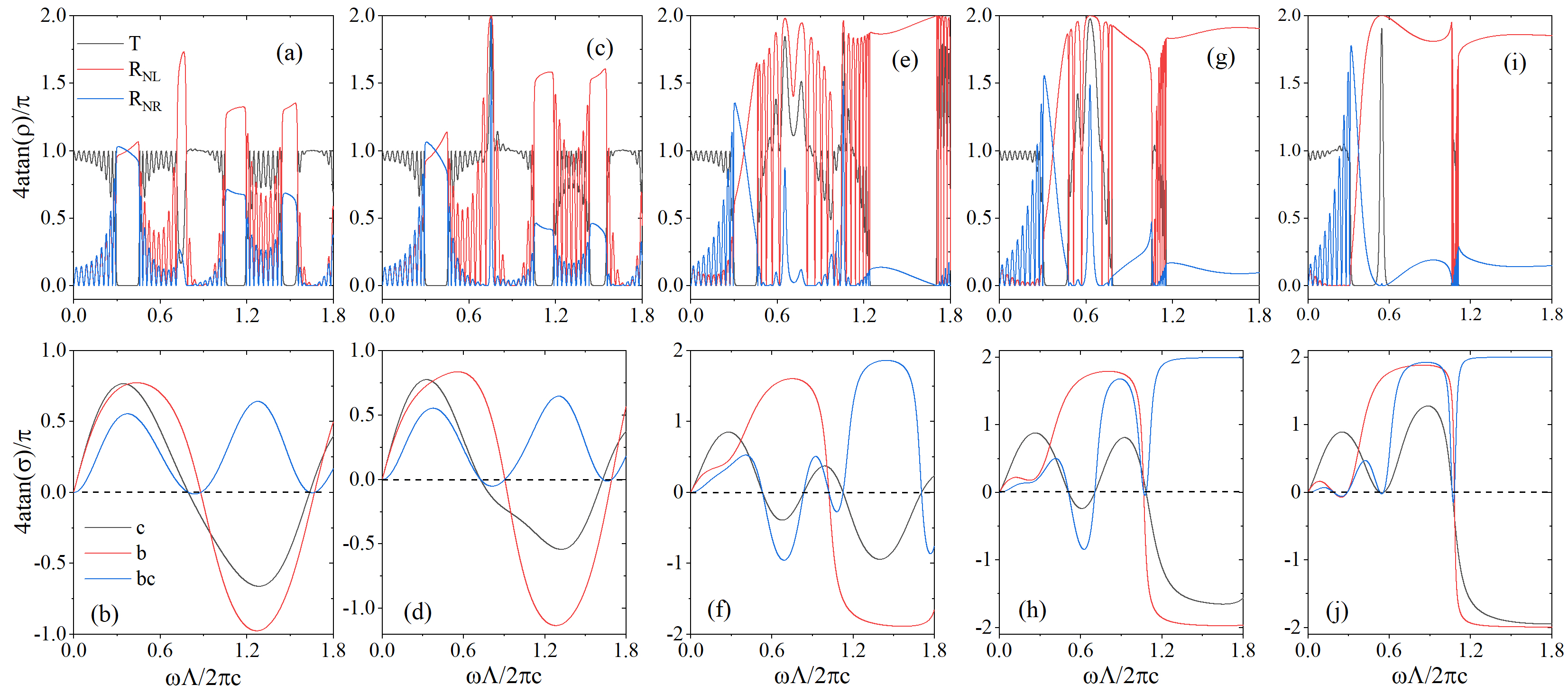}
	
	\caption{\label{fig:5}(Color online) Transmittance ($T_N$) and reflectances ($R_{NL}$ and $R_{NR}$) of the 1DPTSPC with $N=10$ periods. $T_N$, $R_{NL}$ and $R_{NR}$ are denoted by the black, red and blue lines in the top five panels, $c$, $b$ and $bc$ are marked by the black, red and blue lines in the bottom five panels. (a)-(b): $\varepsilon''=0.1$. (c)-(d): $\varepsilon''=0.2$. (e)-(f): $\varepsilon''=1.0$. (g)-(h): $\varepsilon''=1.5$. (i)-(j): $\varepsilon''=2.0$. These curves are also obtained by using transfer matrix method.}
	
\end{figure*}

In Fig.~\ref{fig:4}, we present the complex band structures for different values of the imaginary relative permittivity ${\varepsilon }''$. In our calculation, plane wave expansion method is applied, and the electromagnetic field in the 1DPTSPC is expanded by 61 plane waves. For small values of ${\varepsilon }''$, e.g. ${\varepsilon }''=0.1$ in Figs.~\ref{fig:4}(a) and \ref{fig:4}(b), the eigenfrequencies for all the bands remains real. When ${\varepsilon }''$ is increased to a critical value, the first band gap starts to vanish and the 1st and 2nd bands begin to coalesce, in which case an exceptional point emerges at the center of the Brillouin zone. As ${\varepsilon }''$ is further increased, the previous exceptional point at the center is split in two, e.g. ${\varepsilon }''=0.2$ in Figs.~\ref{fig:4}(c) and \ref{fig:4}(d). There points are located at either sides of the Brillouin center; between these two exceptional points the complex eigenfrequencies emerge, which are conjugate to each other. The band region with complex eigenfrequencies forms the $\mathcal{PT}$-broken phase, while the band region with real eigenfrequencies corresponds to the $\mathcal{PT}$-exact phase. These properties agree with non-Hermitian quantum mechanics.

If ${\varepsilon }''$ is further increased, the exceptional points which originate from the coalescence of the 3rd and 4th bands emerge at the Brillouin edge; at this time, the corresponding $\mathcal{PT}$-broken phase expands, as evident from Figs.~\ref{fig:4}(e) and \ref{fig:4}(f). If we further increase ${\varepsilon }''$ beyond 1.0, the 2nd band gap reopens, and the $\mathcal{PT}$-broken phase expands, which originates from the coalescence of the 2nd and 3rd band, emerges at the Brillouin center and evolves into a band where the eigenvalues are all complex conjugates for all Bloch wave vectors. The $\mathcal{PT}$-broken phase, which originates from the coalescence of the 3rd and 4th bands, emerges at the Brillouin edge and then expands; this expansion then stops at some critical value of ${\varepsilon }''$, after which this $\mathcal{PT}$-broken phase is decreased and finally evolves into a completely real band. These phenomena can be seen in Figs.~\ref{fig:4}(g), \ref{fig:4}(h), \ref{fig:4}(i) and \ref{fig:4}(j). Most of these preceding discussions are disclosed in \cite{w28};  here we investigate further the behaviour of the complex bands after ${\varepsilon }''$ is increased beyond 1.0. We also compare these complex bands with the corresponding real bands, which can be calculated by using transfer matrix method, they are consistent with each other very well.

Figure \ref{fig:5} shows the transmittance, reflectances from both sides and parameters $b$, $c$ and $bc$ for different values of ${\varepsilon }''$. Because the values of the optical properties are very large in some frequency ranges, in order to observe their variation we propose new axis and take the ordinate as ${y=}\frac{4}{\pi }{{{\tan}}^{{-}{1}} \rho}$ for the first time, where $\rho$ denotes the values of these optical properties. In this scale, if $\rho$ is positive (negative), then ${y}$ is also positive (negative). More than that, if $0<\rho<1 (1<\rho<\infty)$, then ${0<y<1 (1<y<2)}$, while $\rho $ tends to ${\pm }{\infty }$, $y$ tends to ${\pm }{2}$. For the same reason, we also take the ordinate as $\frac{4}{\pi }{{{\tan}}^{{-}{1}} \sigma}$ for the parameters we study, where $\sigma$ denotes the values of the parameters.  Combined with Figs.~\ref{fig:4} and \ref{fig:5}, it can be easily seen that these scattering properties are closely related to the band structure. For some conduction bands whose all eigenstates in the complex band structure are in the $\mathcal{PT}$-exact phase, $K{\Lambda}$ varies from $m\pi$ to $(m+1)\pi$. In addition to the premise that $b$ and $c$ are finite, there are also $N-1=9$ points where transmission resonances occur, like those in Fig.~\ref{fig:3}. For ${\varepsilon }'' \ne 0$, a part of a complex conduction band with real frequencies may be not complete, that is to say, the corresponding $K{\Lambda}$ only varies in a part of $\left[m\pi,\ (m+1)\pi \right]$. Consequently, in this band region there are fewer than $N-1=9$ points corresponding to the transmission resonances $T_N=1$, at the same time, $R_{NL}=R_{NR}=0$, these can be also derived from the analytic expressions Eqs.~(\ref{EQ__15_}a)-(\ref{EQ__15_}c). In the band gap, the transmittance is zero, i.e. $T_N=0$, so $\sqrt{R_LR_R}=1$ can be derived from the generalized conservation law, that means, either of $R_{NL}$ and $R_{NR}$ is greater than 1 and the other is less than 1, or both are equal to 1.

As can be seen from Figs.~\ref{fig:4} and \ref{fig:5}, near the position of the exceptional point, where ``near'' means that the frequency or ${\varepsilon }''$ is taken near the corresponding values of the exceptional point, $bc$ becomes negative, so from Eq.~(\ref{EQ__15_}a) the transmittance $T_N$ is larger than 1. Moreover, if $\sin^2 NK\Lambda/\sin^2 K\Lambda$ approaches $\left|1/bc\right|$, singular scattering arises, i.e., the transmittance and reflectances can reach very large values simultaneously from Eqs.~(\ref{EQ__15_}a)-(\ref{EQ__15_}c). For example, when ${\varepsilon }''=0.2$, at $\omega\Lambda/{2\pi c}=0.7535$, $T_N$ reaches 689.35, $R_{NL}$ and $R_{NR}$ reach 8804.24 and 53.82, respectively, as shown in Fig.~\ref{fig:5}(c). Near the exceptional points for ${\varepsilon }''=1.0$ and ${\varepsilon }''=1.5$, we also find that these scattering properties achieve values much larger than one simultaneously. However, in other conduction band regions and all the band gaps, $T_N$, $R_{NL}$ and $R_{NR}$ do not all reach large values simultaneously, because in these regions $bc$ is greater than zero, so, as a result of Eq.(\ref{EQ__15_}a), $T_N$ cannot be great than 1.

In the band gap, as $\varepsilon''$ is increased from 0 to 2.0, the gap between $R_{NL}$ and $R_{NR}$ becomes very large. When ${\varepsilon }''$ is greater than 0.5, we see that $R_{NL}$ is very large and $R_{NR}$ is very small. We refer to the phenomenon where the reflectance from one side is small and that from the other side is large as unidirectional weak visibility. There are some similarities and differences between this phenomenon and unidirectional invisibility. In unidirectional invisibility \cite{w11,w12,w13,w14,w15}, the reflectance from one side is zero and reflectance from the other side is not zero. However, in our phenomenon, the reflectance from one side is also very small but not zero, while the reflectance from the other side is very large. More than that, in unidirectional invisibility the transmittance is one, while in our phenomenon it is zero and occurs in the band gap. This is a significant difference from unidirectional invisibility. For example, $R_{NL}$ is greater than $9.28$ and $R_{NL}$ is less than $0.108$ in the second band gap (1.245 ${<}$ reduced frequency ${<}$ 1.7) for ${\varepsilon }''=1.0$, while $R_{NL}$ is greater than $7.5$ and $R_{NL}$ is less than $0.133$ in the second band gap (1.16 ${<}$ reduced frequency ${<}$ 1.89) for ${\varepsilon }''=1.5$. The unidirectional weak visibility in the band gap is present for a wide frequency range.

\begin{figure}
	
	\centering
	
	\includegraphics[width=.47\textwidth]{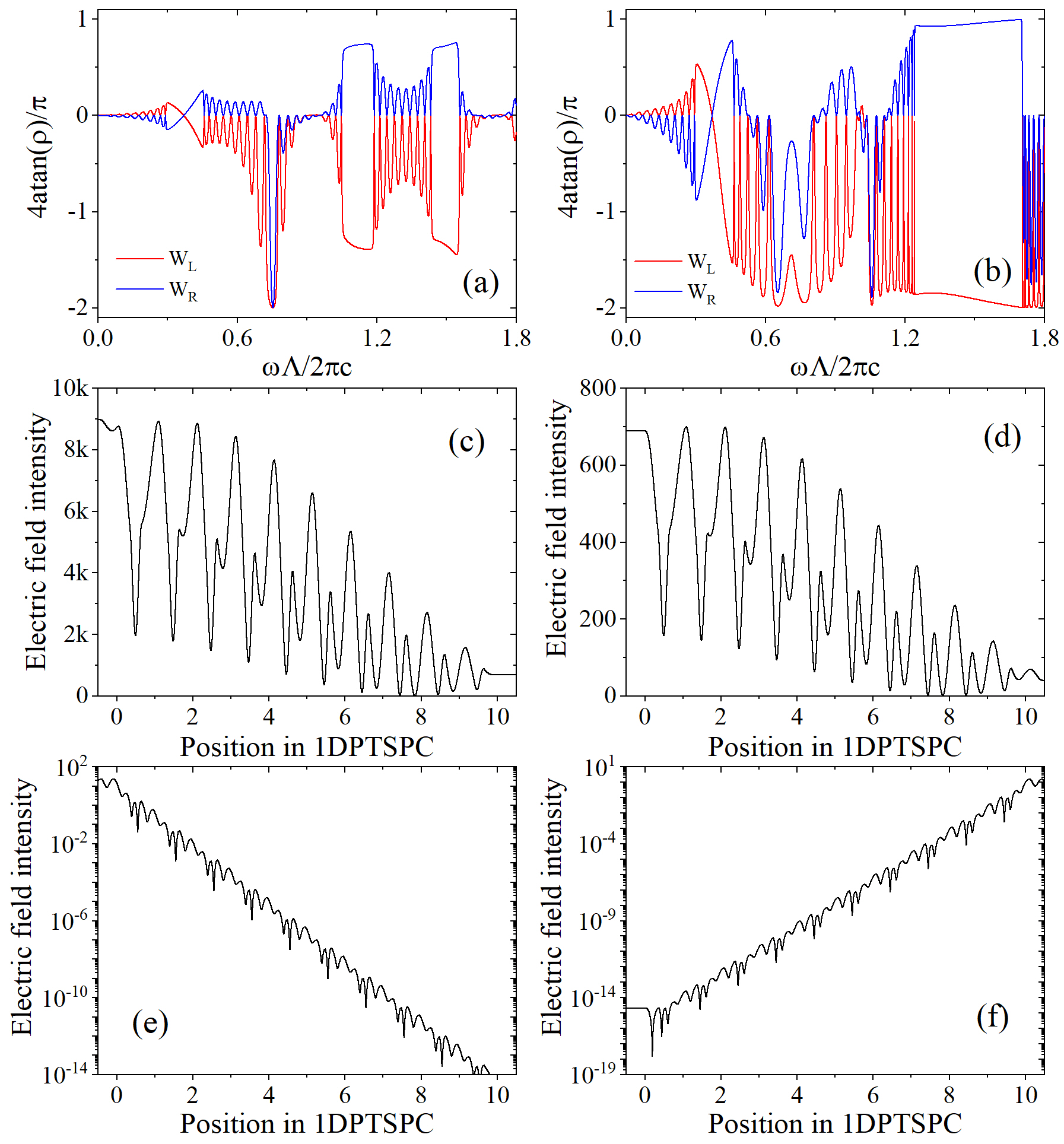}
	
	\caption{\label{fig:6} (Color online) Absorptance spectra ($W_{NL}$ and $W_{NR}$) of the 1DPTSPC with $N=10$ periods at (a) $\varepsilon''=0.2$ and (b) $\varepsilon''=1.0$. Electric field intensity distribution inside the $\mathcal{PT}$-symmetric photonic crystal is shown for incidence from the left side ((c) and (e)) and right side ((d) and (f)). The reduced frequency $\omega\Lambda/{2\pi c}=0.7535$ and $\varepsilon''=0.2$ are chosen for (c) and (d), $\omega\Lambda/2\pi c=1.5$ and $\varepsilon''=1.0$ in a band gap for (e) and (f), respectively. These curves are also obtained by using transfer matrix method.}
	
\end{figure}

In addition to the study on the transmittance and reflectances, we also plot the generalized absorptances with incidence from the left and right sides in Figs.~\ref{fig:6}(a) and \ref{fig:6}(b) when $\varepsilon''=0.2$ and ${\varepsilon }''=1.0$, respectively. It can be seen from Fig.~\ref{fig:6}(a) that near the exceptional point, $W_L$ reaches -9492.6 and $W_R$ reaches -742.2 simultaneously at the singular position, which means the optical effects of the 1DPTSPC from both sides are all dramatic amplification. For the first band gap, the lines of $W_{NL}$ and $W_{NR}$ cross each other, as shown in Figs.~\ref{fig:6}(a) and \ref{fig:6}(b). Actually, this phenomenon also happens for all $\varepsilon''$ we consider. This is also illustrated through Eqs.~(\ref{EQ__17_}d) and (\ref{EQ__17_}e). $bc$ is greater than zero at this region, and $b$ is not equal to $c$ in general; the denominators of $W_{NL}$ and $W_{NR}$ are equal, so if one of their enumerators is more than zero, then the other must be less than zero, so $W_{NL}$ and $W_{NR}$ have different signs. At the cross point $b=c$, so it can be easily seen from Eqs.~(\ref{EQ__17_}d) and (\ref{EQ__17_}e) that $W_{NL}=W_{NR}=0$. In general, when $bc$ is larger than zero, these two absorptances have opposite signs, while when $bc$ is less than zero, these two absorptances are both negative. 

We also investigate the electric field intensity distribution in the 1DPTSPC when the light is incident from the left and right sides. In Figs.~\ref{fig:6}(c) and (d), we plot the intensity distribution when the reduced frequency is chosen at the singular position near the exceptional point for $\varepsilon''=0.2$. It can be easily seen that when the light is incident from the left side, the intensity decays along the propagation direction; from the right side, the intensity grows along the propagation direction. The intensity distribution when the reduced frequency is fixed at the band gap is also considered in Figs.~\ref{fig:6}(e) and \ref{fig:6}(f). We find that the field intensity decays sharply along the propagation direction when light in the band gap is incident from both sides, so the logarithmic coordinate is used to denote the variation. These phenomena agree with our preceding discussions.

\begin{figure*}
	\includegraphics[width=.98\textwidth]{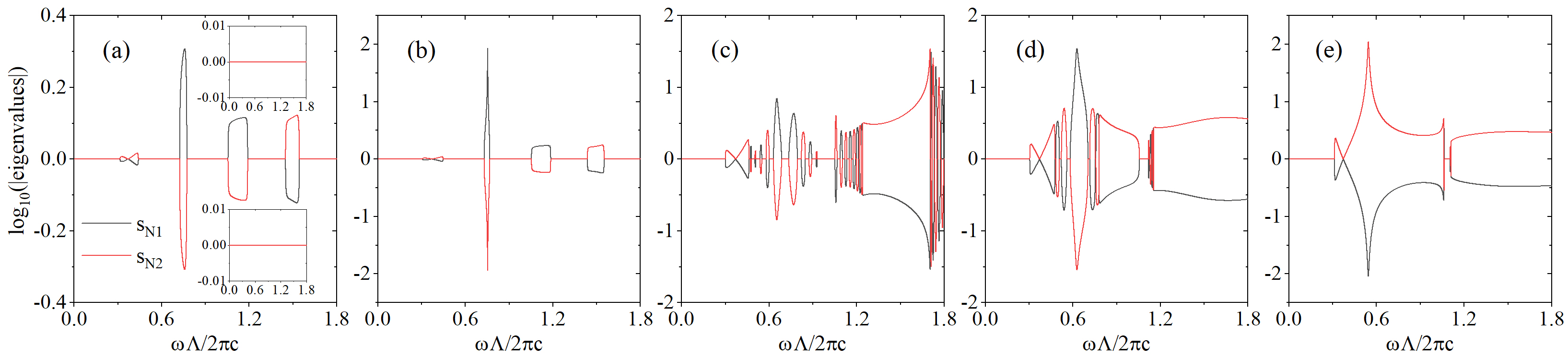}
	
	\caption{\label{fig:7} (Color online) $\text{log}_{10} |s_{N1}|$ (black line) and $\text{log}_{10} |s_{N2}| $ (red line) of the scattering matrix of the 1DPTSPC as a function of the reduced frequency for various values of $\varepsilon''$. (a) $\varepsilon''=0.1$. (b): $\varepsilon''=0.2$. (c) $\varepsilon''=1.0$. (d) $\varepsilon''=1.5$. (e) $\varepsilon''=2.0$. The top inset in (a): $\varepsilon''=0$. The bottom inset in (a): $\varepsilon''=0.001$.}
	
\end{figure*}

\section{\label{sec:level5}$\mathcal{PT}$-SYMMETRY BREAKING DEFINED BY SCATTERING MATRIX}

In the preceding sections, the criterion for differentiating $\mathcal{PT}$-exact and $\mathcal{PT}$-broken phase is originated from Bloch theorem, where periodic boundary condition is applied. Researchers also proposed another criterion \cite{w21,w22,w23}, which is through the eigen-equation of the scattering matrix, this presents more details about the scattering properties of $\mathcal{PT}$-symmetric photonic structure. By this method, in this section we investigate more scattering properties of 1DPTSPCs.

The scattering matrix can be derived from the corresponding total transfer matrix Eq.~(\ref{EQ__12_}), then we can obtain its eigenvalues $s_{N1,N2}$ and the ratios $v_{N1,N2}$ of the two amplitudes of the corresponding normalized eigenstates $\psi_{N1,N2}$

\begin{subequations}
	
	\begin{equation}
		s_{N1,N2}=\frac{i[(b+c)U_N \pm \sqrt{(b-c)^2U_N^2-4}]}{2(aU_N-U_{N-1})},
	\end{equation}
	
	\begin{equation}
		v_{N1,N2}=\frac{i[(c-b)U_N \pm \sqrt{(b-c)^2U_N^2-4}]}{2},
	\end{equation}
	
\end{subequations}

\noindent These two eigenvalues satisfy $\left| s_{N1} s_{N2} \right|=1$. In this criterion, the phase boundary can be derived from the coalesce of these two eigenvalues and eigenstate, which corresponds to the exceptional points. In $\mathcal{PT}$-exact phase, $P \equiv (R_{NL}+R_{NR})/2-T_N-1$ is less than zero, and $\left| s_{N1} \right|=\left| s_{N2} \right|=1$, so the eigenstates show no net amplification nor dissipation. In $\mathcal{PT}$-broken phase, $P$ is greater than zero, and $s_{N1,N2}=1/s_{N2,N1}^*$, hence one eigenstate exhibits amplification and the other exhibits dissipation, and the degree of this amplification or dissipation can be measured by the absolute value of the corresponding eigenvalue.

By using Eqs.~(\ref{EQ__15_}a)-(\ref{EQ__15_}c), when the frequency lies in a conduction band, the index for discriminant $P$ can be obtained

\begin{eqnarray} \label{EQ__31_}
 P = \frac{0.5(b-c)^2-2{{{\sin}^2 K_a {\Lambda}\ }}/{{{\sin}^2 NK_a{\Lambda} }}}{bc+{{{\sin}^2 K_a {\Lambda}\ }}/{{{\sin}^2 NK_a{\Lambda} }}}.
\end{eqnarray}

\noindent On the other hand, when the frequency lies in a band gap, we have

\begin{equation} \label{EQ__32_}
 P = \frac{0.5(b-c)^2-2{{{\sinh}^2 K_{ai} {\Lambda}\ }}/{{{\sinh}^2 NK_{ai}{\Lambda} }}}{bc+{{{\sinh}^2 K_{ai} {\Lambda}\ }}/{{{\sinh}^2 NK_{ai}{\Lambda} }}}.
\end{equation}

\noindent For Eqs.~(\ref{EQ__31_}) and (\ref{EQ__32_}), the denominators are positive, which can be seen from the expressions of the transmittances Eqs.~(\ref{EQ__15_}a) and (\ref{EQ__17_}a), the two terms of each of the two numerators located at either sides of the minus sign are non-negative. In the region of conduction band, there are no definite relation between the values of the corresponding two terms, so either $\mathcal{PT}$-exact or $\mathcal{PT}$-broken phases may arise. In the region of band gap, the second term ${{{\sinh}^2 K_a {\Lambda}\ }}/{{{\sinh}^2 NK_a{\Lambda} }}$ is very small and approaches to zero for large $N$, in general $b\ne c$, then $P>0$, so in this region $\mathcal{PT}$-broken phase arises.

Any input state $\psi$ can be expressed as linear superposition of the eigenstates
\begin{equation}
	\psi=\alpha \psi_{N1}+\beta \psi_{N2},
\end{equation}
 except that $\psi$ is located at the positions of exceptional points where the eigenstates coalesce. The corresponding input intensity is given by
 \begin{equation}
 	|\psi|^2=|\alpha|^2 +2\text{Re}[\alpha^* \beta s_{N1}^* s_{N2} \psi_{N1}^\dagger \psi_{N2}]+|\beta|^2.
 \end{equation}
 In the following, the scattering properties in three different cases ($\mathcal{PT}$-exact phase, $\mathcal{PT}$-broken phase and exceptional points) are expounded for any input state. We will see that amplification, or dissipation, or conservation may be observed for all these cases.

In $\mathcal{PT}$-exact phase, we have the output intensity
\begin{equation}
	\left| S \psi \right|^2-\left| \psi \right|^2=2 \text{Re} \lbrack \alpha^*\beta(s_{N1}^* s_{N2}-1)\psi_{N1}^\dagger\psi_{N2} \rbrack.
\end{equation}
When the input state is either eigenstate ($\alpha=0$ or $\beta=0$) or $b=c$ (for example $\varepsilon''=0$), the intensity is conserved.

In $\mathcal{PT}$-broken phase, it can be shown that
\begin{equation}
	\left| S \psi \right|^2-\left| \psi \right|^2=\alpha^* \alpha (\left|s_{N1} \right|^2-1)+\beta^* \beta (1/\left|s_{N1} \right|^2-1).
\end{equation}
If the amplitudes of $\alpha$ and $\beta$ are equal to each other and their phases are arbitrary, the intensity is amplified.

By this theory, unidirectional weak invisibility can be explained very well. In a band gap, $U_N^2$ is very large for slightly large $N$, consequently, $s_{N1,N2} \approx {i[(b+c)U_N \pm |{(b-c)U_N}|]}/{2(aU_N-U_{N-1})}$. Accordingly, $v_{N1,N2} \approx {i[(c-b)U_N \pm |{(b-c)U_N}|]}/2$, then, one of $v_{N1}$ and $v_{N2}$ approach zero, the other reaches a number with large absolute value. In this case if $(b-c)U_N>0$, the eigenvalues and eigenstates can be written as 
\begin{subequations}
\begin{equation}
s_{N1} \approx {ibU_N}/{(aU_N-U_{N-1})}=r_{NL},
\end{equation}
\begin{equation}
s_{N2} \approx {icU_N}/{(aU_N-U_{N-1})}=r_{NR}, 
\end{equation}
\begin{equation}
\psi_{N1} \approx (\begin{matrix} 1 & 0 \end{matrix})^\text{T}, \quad \psi_{N2} \approx (\begin{matrix} 0 & 1 \end{matrix})^\text{T}.
\end{equation}
\end{subequations}
While if $(b-c)U_N<0$, the situations for $N1$ and $N2$ are reversed. Thereupon, the input beam only from the left side is one eigenstate, while the input beam from the right side is the other eigenstate. In this case, the scattering of one eigenstate exhibits amplification, while the scattering of the other eigenstate exhibits dissipation. As a result, under the condition of zero transmittance, the reflectance from one side can be very large, at the same time the reflectance from the other side can be very small.

In addition to this phenomenon, the singular behavior that transmittance and reflectances from both sides can reach large values simultaneously can be also explained very well. Near the exceptional points achieved by the first criterion in former sections, $bcU_N^2+1$ approaches zero, i.e. $bcU_N^2+1=\delta^2$, where $\delta$ is the positive infinitesimal quantity. Then $(b-c)^2U_N^2-4=(b+c)^2U_N^2-4\delta>0$, so this case belongs to $\mathcal{PT}$-broken phase. Moreover, $s_{N1,N2}={i[(b+c)U_N \pm (|{(b+c)U_N}|-2\delta^2/|{(b+c)U_N}|)]}/(2\delta e^{i\phi})$, here $\phi$ is the phase of $aU_N-U_{N-1}$. Correspondingly, we have the ratios $v_{N1,N2}={i[(c-b)U_N \pm (|{(b+c)U_N}|-2\delta^2/|{(b+c)U_N}|)]}/2$. In this case, if $(b+c)U_N>0$, the eigenvalues and ratios for the eigenstates are given by

\begin{subequations}
	\begin{equation}
		s_{N1}=\frac{i(b+c)U_N-i\delta^2/[(b+c)U_N]}{\delta e^{i \phi}},
	\end{equation}
	\begin{equation}
		s_{N2}=\frac{i\delta^2/[(b+c)U_N]}{\delta e^{i\phi}},
	\end{equation}
	\begin{equation}
		r_{N1}=icU_N, \quad r_{N2}=-ibU_N.
	\end{equation}
\end{subequations}
Then it can be derived that $|s_{N1}| \to \infty$ and $|s_{N2}| \to 0$ when $\delta \to 0$, i.e. the pole and zero of the scattering matrix can be approached. While if $(b+c)U_N<0$, the situations for $N1$ and $N2$ are reversed. For example, when $\omega\Lambda/{2\pi c}=0.7535$ for $\varepsilon''=0.2$, $\delta^2$ reaches the minimum $1.45 \times 10^{-3}$, so singular behavior appears here.

\begin{figure}
	\includegraphics[width=.45\textwidth]{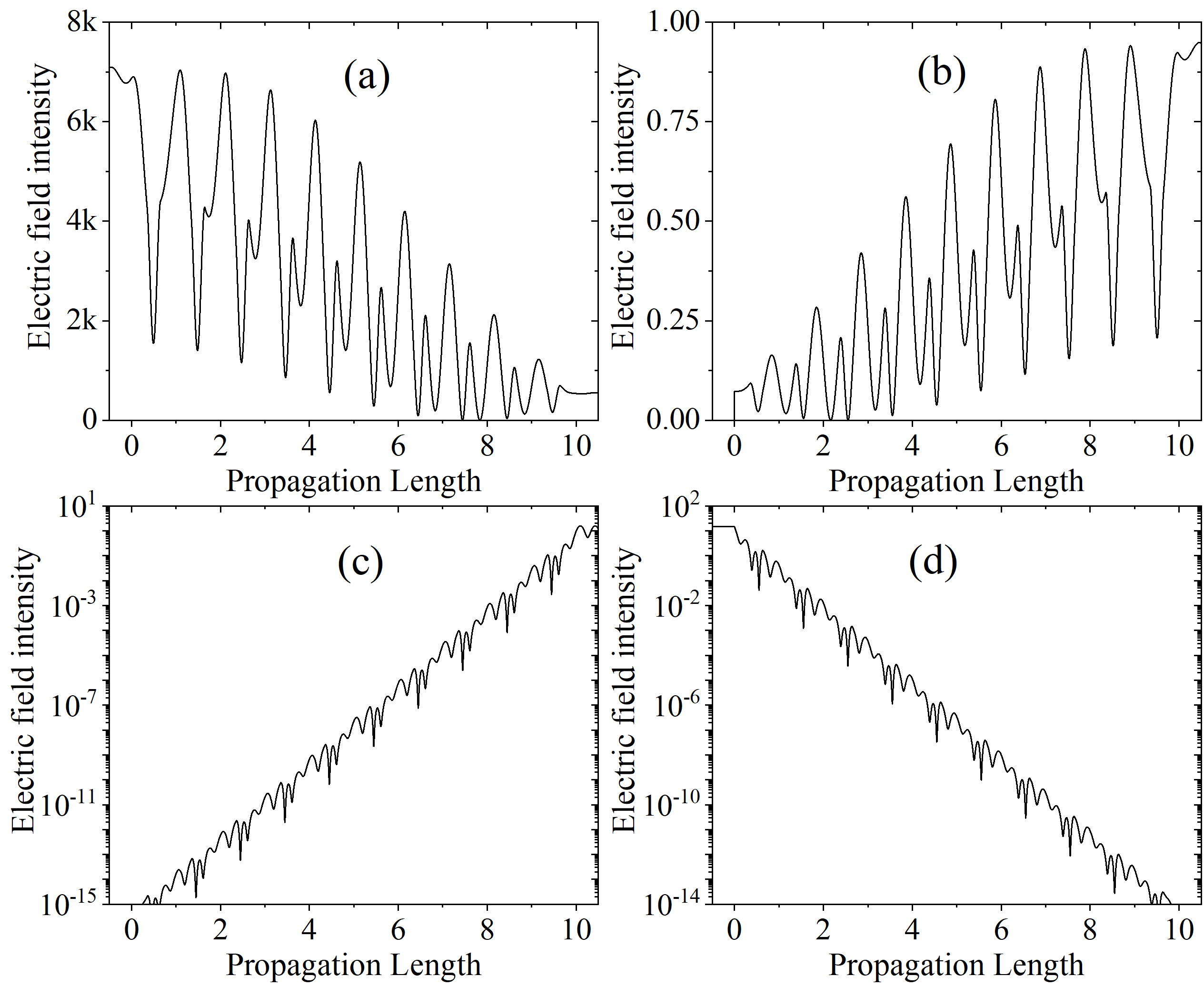}
	
	\caption{\label{fig:8} Electric field distribution of scattering eigenstates $\psi_1$ ((a) and (c)) and $\psi_2$ ((b) and (d)). The reduced frequency $\omega\Lambda/{2\pi c}=0.7535$ and $\varepsilon''=0.2$ are chosen for (a) and (b), where $|s_1|=86.4782$ and $|s_2|=0.0116$; $\omega\Lambda/2\pi c=1.5$ and $\varepsilon''=1.0$ in a band gap are assumed for (c) and (d), where $|s_1|=0.2586$ and $|s_2|=3.8664$.}
	
\end{figure}

At the exceptional points defined by the criterion in present section, where $(b-c)U_{N-1}=\pm 2$, the eigenvalues $s_{1,2}$ and eigenvectors $\psi_{1,2}$ coalesce.
\begin{subequations}
	
	\begin{equation}
		s_{N1}=s_{N2} \equiv s_N=\frac{i(b U_N \mp 1)}{aU_N-U_{N-1}},
	\end{equation}
	
	\begin{equation}
		\psi_{N1}=\psi_{N2} \equiv \psi_{Na}=\frac{1}{\sqrt{2}} \left( \begin{array}{cc} 1 \\ \mp i \end{array} \right).  
	\end{equation}
	
\end{subequations}
\noindent In order to express any input state at the exceptional points as linear superposition of the eigenstate, the auxiliary state $\psi_{Nb}$ is introduced that is orthogonal to $\psi_{Na}$
\begin{equation}
\psi_{Nb}=\frac{1}{\sqrt{2}} \left( \begin{array}{cc} 1 \\ \pm i \end{array} \right). 
\end{equation}
Similarly, any input state at the exceptional point can be written as
\begin{equation}
	\psi=\alpha \psi_{Na}+\beta \psi_{Nb}.
\end{equation}
It can be derived that
\begin{equation}
	\left| S \psi \right|^2-\left| \psi \right|^2=\frac{4 \text{Re}(\alpha^* \beta)}{\pm bU_N-1}+\frac{4 \left| \beta \right|^2}{\left| aU_N-U_{N-1} \right|^2}.
\end{equation}
If the input state is $\alpha\psi_{Na}$, then the intensity is conserved; otherwise if the input state is $\beta\psi_{Nb}$, then the intensity is amplified.

\begin{figure}
	\includegraphics[width=.48\textwidth]{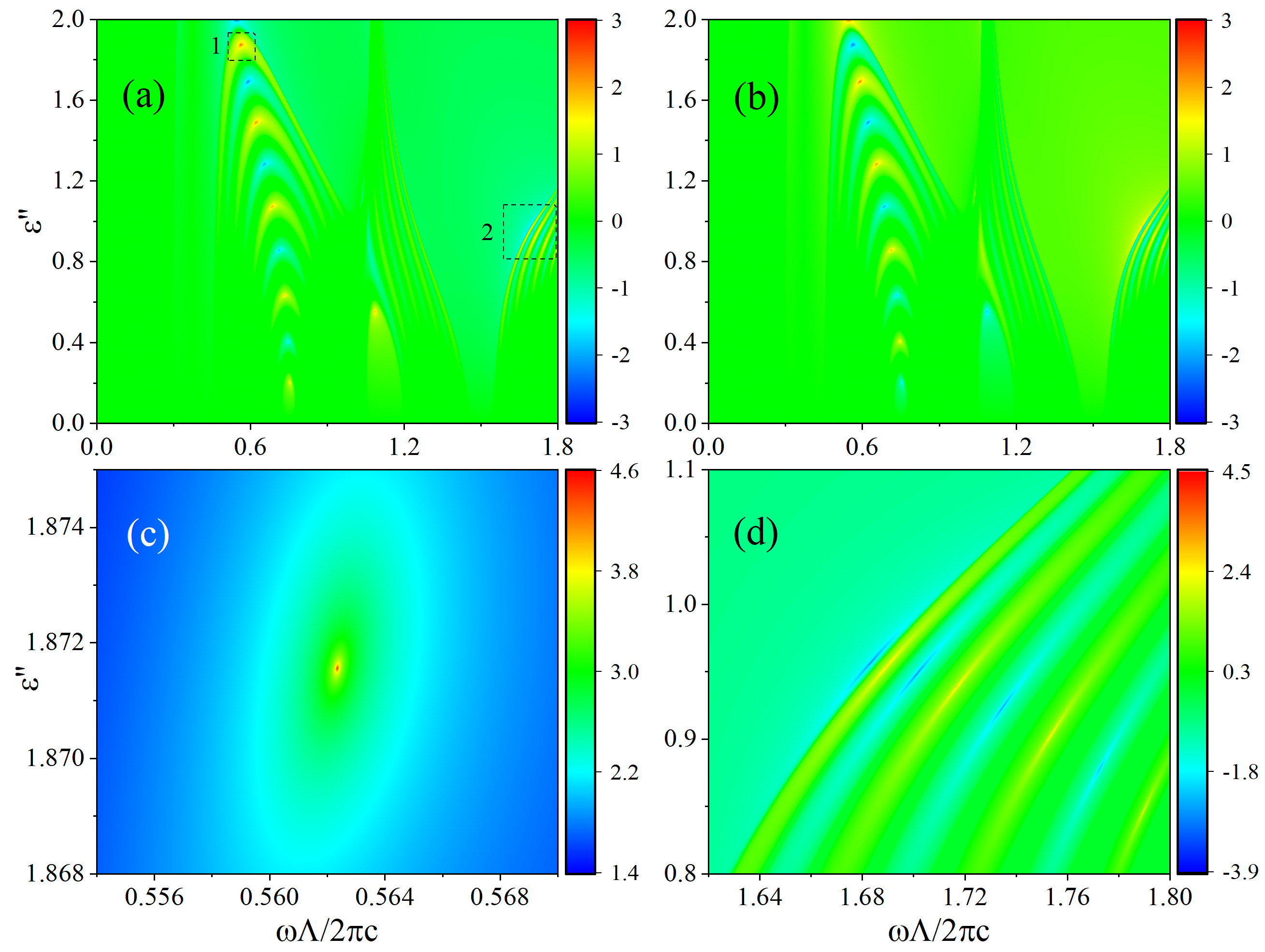}
	
	\caption{\label{fig:x} (Color online) Eigenvalues $\text{log}_{10} |s_1|$ (a) and $\text{log}_{10} |s_2| $ (b) of the scattering matrix of the 1DPTSPC as functions of the reduced frequency $\omega\Lambda/{2\pi c}$ and imaginary part of permittivity $\varepsilon''$. (c) Magnified map near a pole in the square region 1 of (a). (d) Magnified map for the square region 2 in (a).}
	
\end{figure}

Figure \ref{fig:7} illustrates the modulus of the eigenvalues $s_{N1}$ and $s_{N2}$ for different values of $\varepsilon''$. In absence of gain and loss, as shown in the top inset of Fig.~\ref{fig:7}(a), $\left| s_{N1} \right|=\left| s_{N2} \right|=1$ throughout the spectrum. In this case, $b=c$ out of $\mathcal{T}$-symmetry, then $P<0$ derived from Eqs. (\ref{EQ__31_}) and (\ref{EQ__32_}), so the system is in $\mathcal{PT}$-exact phase. When $\varepsilon''$ is increased slightly, for example, $\varepsilon''=0.001$ in the bottom inset, $\mathcal{PT}$-exact phase is also found throughout the spectrum. As $\varepsilon''$ is increased continuously, $\mathcal{PT}$-broken phase arises in the band gaps and some parts of the conduction bands, which agrees with our preceding theory.

For $\mathcal{PT}$-broken phase, the electric field intensity distributions of eigenstates in a conduction and a band gap are shown in Figs.~\ref{fig:8}(a), \ref{fig:8}(b) and Figs.~\ref{fig:8}(c), \ref{fig:8}(d), respectively. For comparison, the reduced frequencies are identical to those in Fig.~\ref{fig:6}. It can be obviously seen that the field distributions in Figs.~\ref{fig:8}(c) and \ref{fig:8}(d) are the same as those in Figs.~\ref{fig:6}(f) and \ref{fig:6}(e). This is because the input beam from left side is the eigenstate $\psi_{N2}$ while the input beams from right side is the eigenstate $\psi_{N1}$, which is explicated in preceding theory. Moreover, based on this property, 1DPTSPCs can be act as a good absorber and a laser simultaneously. Compared with the system in \cite{w21} which can act simultaneously as a coherent perfect absorber and laser at the pole and zero of the scattering matrix, the effects of absorbing and lasing in the band gap of our system can be easily realized and possess large frequency region.

In order to study the eigen-equation of the scattering matrix comprehensively and thoroughly, we plot the eigenvalues as functions of the reduced frequency and imaginary part of permittivity in Figs.~\ref{fig:x}(a) and \ref{fig:x}(b). We find that the poles and zeros of the scattering matrix are distributed discretely, and they turn up alternately, this is because the sign of $(b+c)U_N$ is changed. The values of eigenvalues are not $0$ or $\infty$, because in our calculation the points we scan are finite. So if we magnify some poles and zeros (for example, in Figs.~\ref{fig:x}(c) and \ref{fig:x}(d)) we find the corresponding values at the poles and zeros are increased and decreased, respectively. These phenomena conform to our theory, by our calculation, at these positions $bcU_N^2+1=0$. 

\begin{figure}
	\includegraphics[width=.49\textwidth]{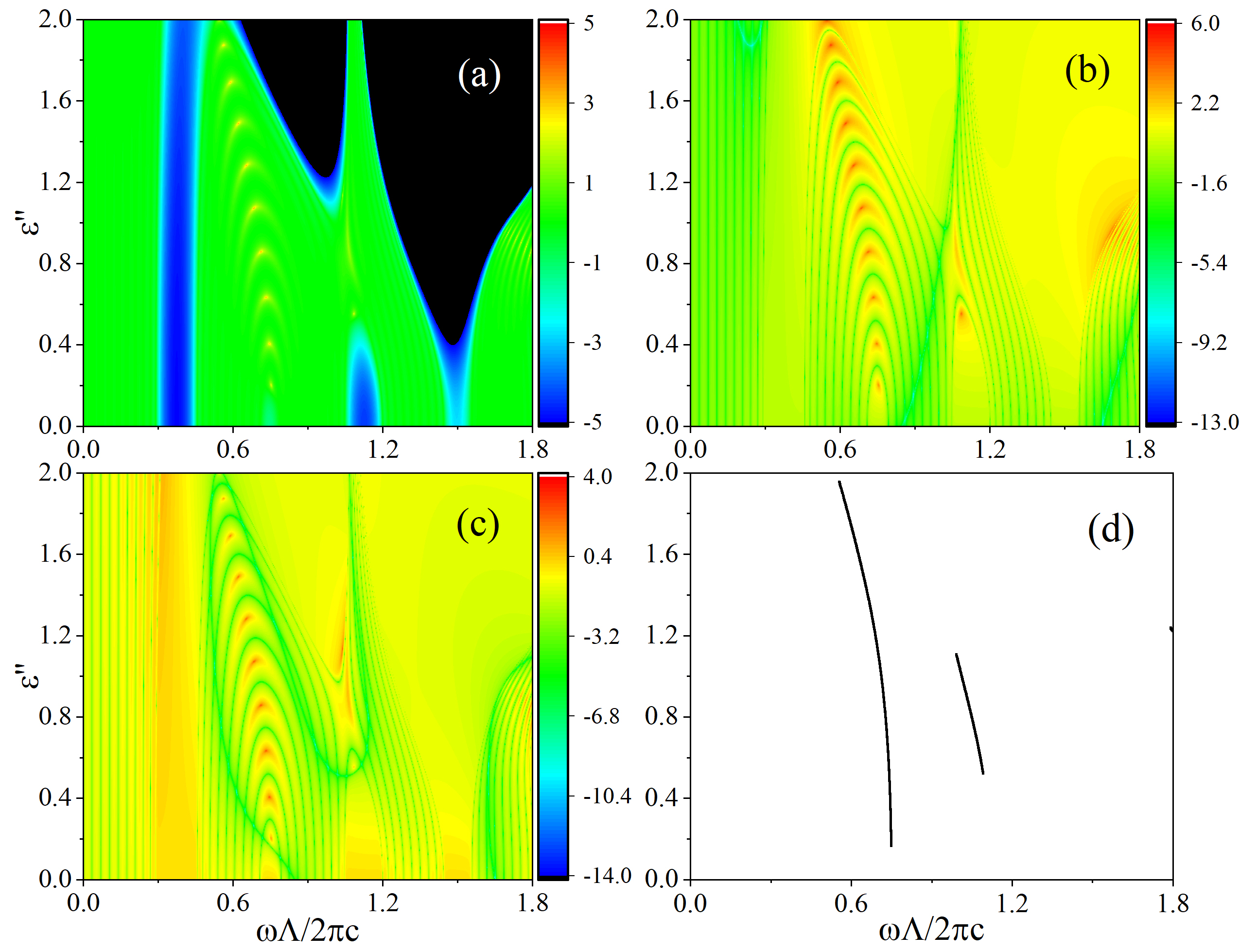}
	
	\caption{\label{fig:y} (Color online) (a) $\text{log}_{10}(T_{N})$, (b) $\text{log}_{10}(R_{NL})$ and (c) $\text{log}_{10}(R_{NR})$ as functions of the reduced frequency $\omega\Lambda/{2\pi c}$ and imaginary part of permittivity $\varepsilon''$. (d) The trajectory of the exceptional points in the same space $(\omega\Lambda/{2\pi c}, \varepsilon'')$, which is obtained from the complex band structures in the former sections.}
	
\end{figure}

\begin{figure}
	\includegraphics[width=.49\textwidth]{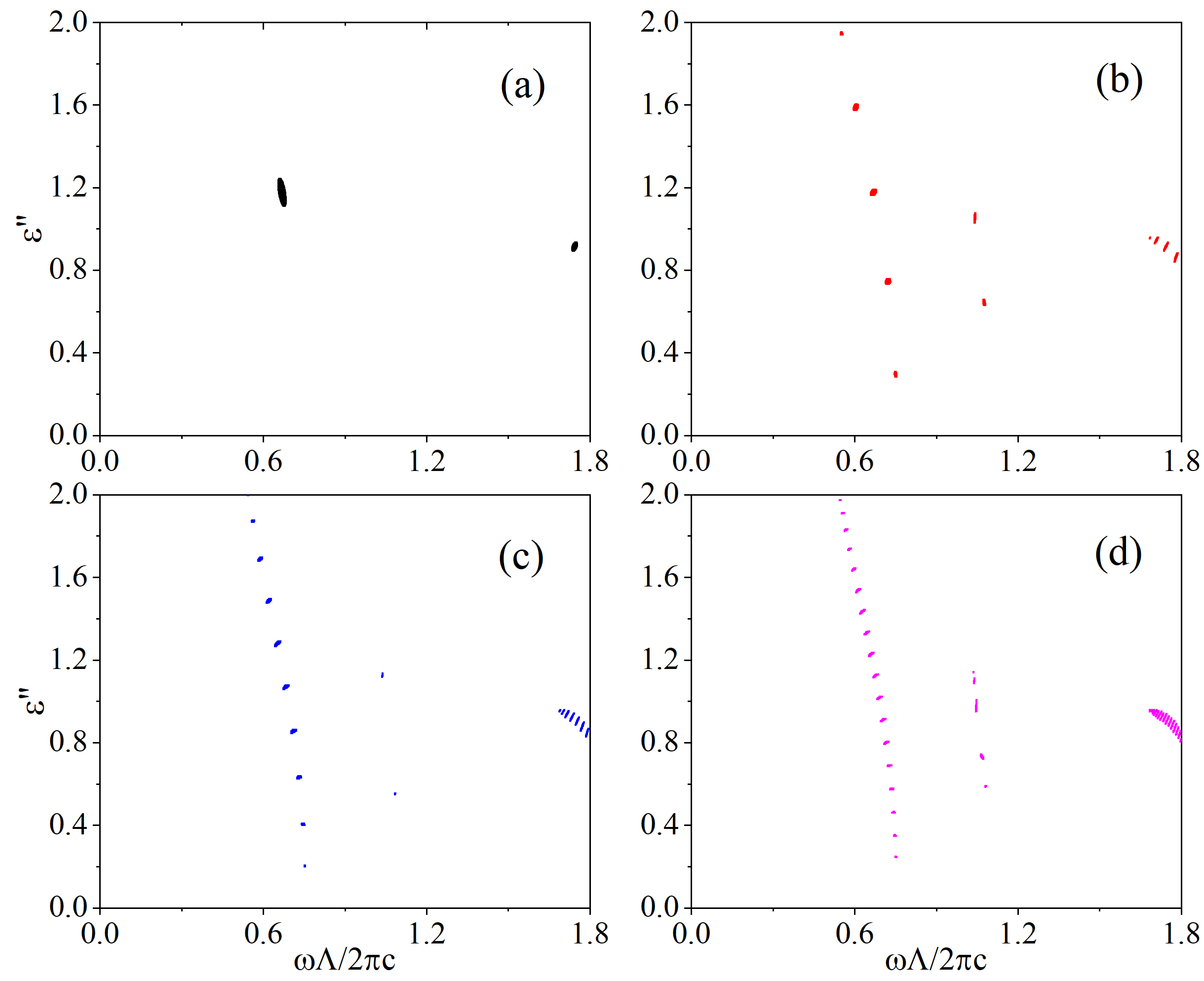}
	
	\caption{\label{fig:z} (Color online) The points of singular scattering in the space $(\omega\Lambda/{2\pi c}, \varepsilon'')$ with $N=1$ (a), $N=5$ (b), $N=10$ (c) and $N=20$ (d).}
	
\end{figure}

The relation between the poles and zeros and the singular scattering can be seen clearly in Figs.~\ref{fig:y}(a), \ref{fig:y}(b) and \ref{fig:y}(c), respectively, where $\text{log}_{10}(T_{N})$, $\text{log}_{10}(R_{NL})$ and $\text{log}_{10}(R_{NR})$ in the same space $(\omega\Lambda/{2\pi c}, \varepsilon'')$ are depicted. It can be illustrated that at the poles and zeros of the scattering matrix singular scattering appears. Moreover, the black and blue regions in Fig.~\ref{fig:y}(a) are band gaps with zero transmittance, here the unidirectional weak visibility happens, which also can be also seen in Figs.~\ref{fig:y}(b) and \ref{fig:y}(c).

Figure~\ref{fig:y}(d) shows the trajectory of the exceptional points obtained from the complex band structure in the first criterion in the space $(\omega\Lambda/{2\pi c}, \varepsilon'')$, while Fig.~\ref{fig:z} present the points of singular scattering in the same space. Comparing these figures, we see that the points of singular scattering are distributed discretely close to the trajectory of the exceptional points. When the number of unit cells $N$ is increased, more points of singular scattering arise. This phenomenon can be also explained by our theory, $b$ and $c$ are independent of $N$ and there are more points of $U_N^2$ along the trajectory satisfying $bcU_N^2+1=0$ for large $N$, which can be understood from the function diagram of $U_N^2$. This means that singular scattering can be realized easier for more unit cells.

\section{\label{sec:level6}Relation between two criteria}

The universally accepted criterion in $\mathcal{PT}$-symmetric quantum mechanics \cite{w70,w71} is based on the eigen-equation of non-Hermitian Hamiltonian $\hat{H} \Psi_n = \lambda_n \Psi_n$, where $\hat{H}=\hat{p}^2/(2m)+V(x)$, and $[\mathcal{PT},\hat{H}]$ holds because the system we consider is $\mathcal{PT}$-symmetric. Hence the following important relation can be derived $\hat{H} \mathcal{PT} \Psi_n = \lambda_n^* \mathcal{PT} \Psi_n$. If $\mathcal{PT} \Psi_n \propto \Psi_n$, the eigenvalue is real $\lambda_n = \lambda_n^*$, this case is referred to as $\mathcal{PT}$-exact phase, otherwise the eigenvalues $\lambda_n$ and $\lambda_n^*$ form a complex conjugate pair, this case corresponds to $\mathcal{PT}$-broken phase.

In the former sections, two criteria are introduced to define and differentiate $\mathcal{PT}$-exact and $\mathcal{PT}$-broken phases for 1DPTSPCs, the first criterion is based on the eigen-equation of effective non-Hermitian Hamiltonian, the second criterion is based on the eigen-equation of scattering matrix. The first criterion is consistent with the universally accepted criterion for the following reasons. The analog in one-dimensional photonic crystal of the eigen-equaiton of Hamiltonian in quantum mechanics can be obtained from master equation $\hat{H} \Phi_n = \omega_n^2/c^2 \Phi_n$, here $\hat{H}=-1/\sqrt{\varepsilon (x)} \partial^2/\partial x^2  1/\sqrt{\varepsilon (x)}$ and $\Phi_n = \sqrt{\varepsilon(x)}E_n(x)$, where $E_n(x)$ is electric field distribution. After a similar derivation with that in the quantum mechanics, we can define $\mathcal{PT}$-exact phase where the eigenvalue $\omega_n$ is real $\omega_n = \omega_n^*$ and $\mathcal{PT}$-broken phase where the eigenvalues $\omega_n$ and $\omega_n^*$ form a complex conjugate pair. These results are reflected in the complex band structures.

On the other hand, the second criterion is obtained based on the eigen-equaiton of scattering matrix \cite{w21} $ S (\omega) \psi_{Nn} = s_{Nn} \psi_{Nn} (n=1, 2)$. Given $\mathcal{PT}$-symmetry, $ S (\omega^*) \mathcal{PT} \psi_{Nn} = 1/s_{Nn}^* \mathcal{PT} \psi_{Nn}$. When $\mathcal{PT}\psi_{N1,N2} \propto \psi_{N1,N2}$ the eigenstates show no net amplification nor dissipation ($\left|s_{N1,N2}\right|=1$) and this case is defined as $\mathcal{PT}$-exact phase, on the other hand, when $\mathcal{PT}\psi_{N1,N2} \propto \psi_{N2,N1}$ one eigenstate exhibits amplification and the other exhibits dissipation ($s_{N1,N2}=1/s^*_{N2,N1}$), this case is taken as $\mathcal{PT}$-broken phase. The prerequisite for this definition is that the frequency is real. As we can see in $\mathcal{PT}$-exact phase these two eigenvalues $s_{N1,N2}$ are not real and in $\mathcal{PT}$-broken phase these two eigenvalues $s_{N1,N2}$ are not complex conjugate pair, so the second criterion is not consistent with the universally accepted criterion. Hence, these definitions of $\mathcal{PT}$-exact and $\mathcal{PT}$-broken phases are inappropriate. However, these relations provide the important knowledge of the scattering properties for $\mathcal{PT}$-symmetric photonic structure.

According to the above definitions, we can conclude that these two criteria give two different definitions of $\mathcal{PT}$ symmetry breaking and describe different optical properties for $\mathcal{PT}$-symmetric photonic structure. Consequently, the exceptional points defined in these two criterion are different. Therefore, comparing Figs.~\ref{fig:4} and \ref{fig:7}, we can see that the regions of $\mathcal{PT}$-exact phase and $\mathcal{PT}$-broken phase are different and so are exceptional points. For example, exceptional points gotten from effective Hamiltonian in the first criterion is independent of the number of unit cells $N$ because of Bloch theorem, while the exceptional points gotten from the scattering matrix is dependent of $N$. Actually, the optical properties described in the second criterion resides in the $\mathcal{PT}$-exact phase and band gaps of the first criterion, because the frequency in the second criterion is real. The relation between these two criterion is disclosed detailedly in section \ref{sec:level5}.

In reference \cite{w21}, the corresponding non-Hermitian cavity Hamiltonian spectrum of $\mathcal{PT}$-symmetric heterostructure, which corresponds to one unit cell in our work, is derived from the vanished amplitude boundary condition. Actually, this spectrum is composed of some points of whole band structure derived from Bloch theorem. Because in that reference from vanished amplitude boundary condition, we have $B=-A$ and $D=-C$, then the corresponding Bloch wave vector can be obtained $K \Lambda =-i \text{log}(C/A)=-i \text{log}(D/B)$. When the number of unit cells $N$ is increased, this explanation is inaccurate, and the explanation of complex band structure should be introduced.

\section{\label{sec:level7}CONCLUSION}

We have defined the generalized absorptance to include the amplification effect of the active dielectric, derived the mathematical expressions of the scattering properties of the 1DPTSPCs, and expounded and analyzed these scattering properties using two criteria. We find that singular scattering appears close to the exceptional points in conduction bands, namely that the transmittance and reflectances from both sides reach large values simultaneously and approach to infinity, while dramatic amplification from both sides arises. The positions of this singular scattering coincide with the poles and zeros of the scattering matrix. In the band gap, we observe unidirectional weak visibility, as the transmittance is zero and the reflectance is very large from one side, while the reflectance from the other side is very small. Actually, this case corresponds to the scattering of the eigenstates of the scattering matrix. Based on this study, the criterion on the basis of Bloch theorem, which is related to the eigen-equation of effective non-Hermitian Hamiltonian, is preferable. We envision various applications in the designs of functional optical devices \cite{w18,w19,w20,w57,w55}, such as filters, switches, lasers, absorbers, sensors and so on.

\begin{acknowledgements}
This work is supported by the National Natural Science Foundation of China (Grant No. 12004231), the Science and Technology Innovation Planning Project in Universities and Colleges of Shanxi Province of China (Grant No. 2019L0019), and the Applied Basic Research Program of Shanxi Province of China (Grant No. 201901D211165).
\end{acknowledgements}

\bibliography{wtc}

\end{document}